\documentclass[%
 reprint,
superscriptaddress,
nofootinbib,
 amsmath,amssymb,
 aps,
 prc,
]{revtex4-2}

\usepackage{graphicx}
\usepackage{dcolumn}
\usepackage{bm}
\usepackage[colorlinks=true, urlcolor=blue, linkcolor=blue, citecolor=blue]{hyperref}
\usepackage{comment}
\usepackage{gensymb}
\usepackage{xfrac} 
\usepackage{dcolumn}
\usepackage{multirow}
\usepackage{booktabs}
\usepackage{xcolor}
\usepackage{orcidlink}

\begin{document}

\preprint{APS/123-QED}

\title{Ultra-High precision Compton polarimetry at 2\,GeV}

\author{A.~Zec\,\orcidlink{0000-0003-2207-5487}}\affiliation{Department of Physics, University  of  Virginia,  Charlottesville,  Virginia  22904,  USA}
\author{S.~Premathilake}\affiliation{Department of Physics, University  of  Virginia,  Charlottesville,  Virginia  22904,  USA}
\author{J.C.~Cornejo}\affiliation{Physics Department, Carnegie Mellon University, Pittsburgh, Pennsylvania  15213, USA}
\author{M. M.~Dalton\,\orcidlink{0000-0001-9204-7559}}\email{dalton@jlab.org}\affiliation{Thomas Jefferson National Accelerator Facility, Newport News, Virginia 23606, USA}
\author{C.~Gal\,\orcidlink{0000-0003-0076-2120}}\affiliation{Department of Physics, University  of  Virginia,  Charlottesville,  Virginia  22904,  USA}\affiliation{Thomas Jefferson National Accelerator Facility, Newport News, Virginia 23606, USA}\affiliation{Department of Physics and Astronomy, State University of New York, Stony Brook, New York 11794, USA}\affiliation{Center for Frontiers in Nuclear Science, State University of New York, Stony Brook, New York 11794, USA}
\author{D.~Gaskell\,\orcidlink{0000-0001-5463-4867}}\affiliation{Thomas Jefferson National Accelerator Facility, Newport News, Virginia 23606, USA} 
\author{M.~Gericke\,\orcidlink{0000-0002-8976-8192}}\affiliation{Department of Physics and Astronomy, University of Manitoba, Winnipeg, Manitoba R3T2N2 Canada}
\author{I.~Halilovic}\affiliation{Department of Physics and Astronomy, University of Manitoba, Winnipeg, Manitoba R3T2N2 Canada}
\author{H.~Liu\,\orcidlink{0000-0002-5555-9632}}\affiliation{Department of Physics, University of Massachusetts, Amherst, Massachusetts 01003, USA}
\author{J.~Mammei}\affiliation{Department of Physics and Astronomy, University of Manitoba, Winnipeg, Manitoba R3T2N2 Canada}
\author{R.~Michaels}\affiliation{Thomas Jefferson National Accelerator Facility, Newport News, Virginia 23606, USA}
\author{C.~Palatchi}\affiliation{Department of Physics, University  of  Virginia,  Charlottesville,  Virginia  22904,  USA}\affiliation{Center for Frontiers in Nuclear Science, State University of New York, Stony Brook, New York 11794, USA}
\author{J.~Pan\,\orcidlink{0009-0005-7888-7232}}\affiliation{Department of Physics and Astronomy, University of Manitoba, Winnipeg, Manitoba R3T2N2 Canada}
\author{K.D.~Paschke\,\orcidlink{0000-0001-8794-8221}}\affiliation{Department of Physics, University  of  Virginia,  Charlottesville,  Virginia  22904,  USA}
\author{B.~Quinn\,\orcidlink{0000-0003-2800-986X}}\affiliation{Physics Department, Carnegie Mellon University, Pittsburgh, Pennsylvania  15213, USA}
\author{J.~Zhang\,\orcidlink{0000-0002-4478-1289}}\affiliation{Department of Physics and Astronomy, State University of New York, Stony Brook, New York 11794, USA}\affiliation{Center for Frontiers in Nuclear Science, State University of New York, Stony Brook, New York 11794, USA}\affiliation{Institute of Frontier and Interdisciplinary Science and Key Laboratory of Particle Physics and Particle Irradiation (MOE), Shandong University, Qingdao, Shandong 266237, China}

\date{\today}

\begin{abstract}
We report a high precision measurement of electron beam polarization using Compton polarimetry. The measurement was made in experimental Hall A at Jefferson Lab during the CREX experiment in 2020.  A total uncertainty of $dP/P=0.36$\% was achieved detecting the back-scattered photons from the Compton scattering process.  This is the highest accuracy in a measurement of electron beam polarization using Compton scattering ever reported, surpassing the ground-breaking measurement from the SLAC Large Detector (SLD) Compton polarimeter.  Such uncertainty reaches the level required for the future flagship measurements to be made by the MOLLER and SoLID experiments.

\end{abstract}

\maketitle

\section{Introduction}

The Calcium Radius Experiment (CREX) is a precision determination of the neutral weak form factor of $^{48}$Ca~\cite{CREX:2022kgg}.  The form factor is determined from a precise measurement of the parity-violating (PV) asymmetry $A_{\rm PV}$ in elastic scattering of longitudinally polarized electrons from $^{48}$Ca.

The asymmetry of approximately 2.7 ppm was measured to 4\% statistical and 1.5\% systematic uncertainties.  From this asymmetry, the weak form factor and the difference between the weak and charged form factors were extracted.  The resulting neutron skin thickness, with additional uncertainty from the extraction model, is relatively thin yet consistent with many model calculations. 

While the dominant uncertainty in the CREX measurement was  statistical, one of the more important systematic uncertainties was due to the measurement of the beam polarization.  
In many experiments, the leading source of systematic uncertainty is knowledge of the beam polarization~\cite{Abrahamyan:2012gp,PVDIS:2014cmd}.

Polarimetry of an electron beam with GeV energy is accomplished using either Compton scattering from circularly polarized laser photons or M\o ller scattering from  atomic electrons in a polarized metallic foil, where the target photons or electrons must have a known polarization.  
The Compton technique allows continuous monitoring at high electron beam currents synchronous with the experiment while the M\o ller technique samples at specific times with low beam current during which the experiment cannot run.
The electron polarization does not depend significantly on the beam current~\cite{Magee:2016xqx}, but to meet stringent uncertainty goals, parity violation measurements make use of both Compton polarimetry, to continuously monitor variations of beam polarization with time, and M\o ller polarimetry, as it has independent systematic uncertainties with comparable accuracy.

In this paper we focus on the improvements in the Compton polarimetry technique that were made in experimental Hall A at Thomas Jefferson National Accelerator Facility (Jefferson Lab) that resulted in achieving a new level of systematic uncertainty.
In Sec.~\ref{sec:compton} we give an overview of the Compton polarimeter and review the evolution of the system since its initial commissioning more than 20 years ago.
In Sec.~\ref{sec:laser} we describe the setup of the laser which constitutes the polarized photon target and our determination of the laser polarization.
In Sec.~\ref{sec:gso} we describe the system for detecting the high energy scattered photons and its use to determine the electron beam polarization.
In Sec.~\ref{sec:results} we present the results of the electron beam polarization measurement and summarize the uncertainties.

\section{Hall A Compton polarimeter} \label{sec:compton}

\begin{figure*}[htb]
\begin{center}
{\includegraphics*[width=0.75\textwidth]{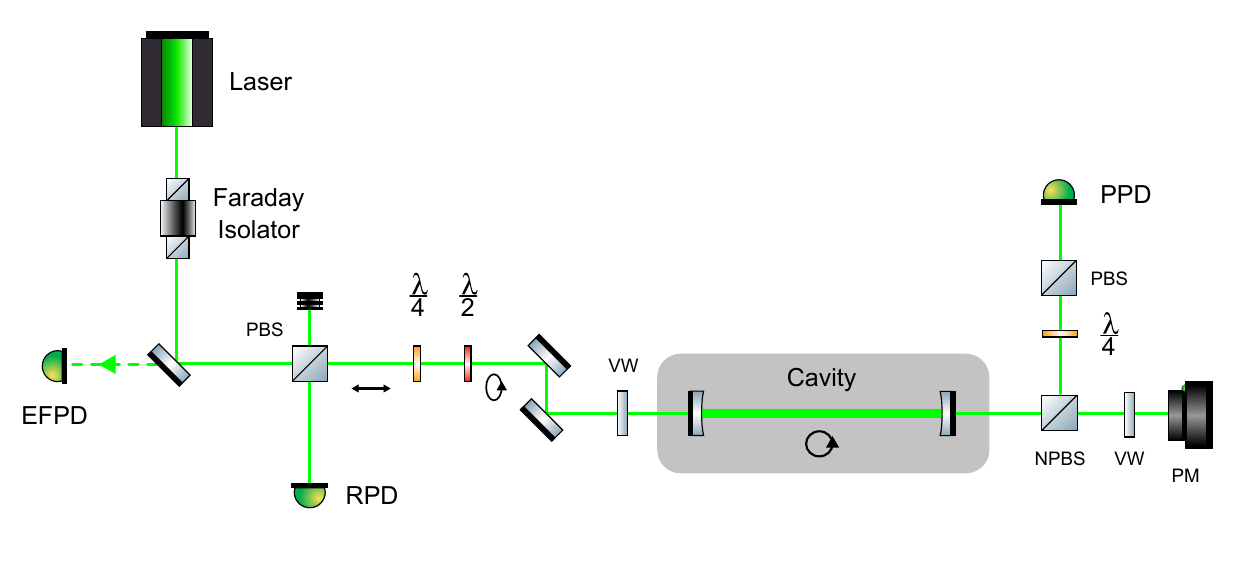}}
\caption{Optics layout for measurement of the degree of circular polarization (DOCP) in the Fabry-P\'{e}rot cavity.  After the first polarizing beam splitter (PBS) the laser polarization is linear, transformed to an arbitrary state by the quarter and half wave plates ($\lambda$/4 and $\lambda$/2) .  The wave plates are typically set so that the polarization will be circular when stored inside the cavity, incorporating the birefringence of the steering mirrors and vacuum window (VW).  The ``entrance function" is determined via measurements in the entrance function photo-diode (EFPD) with the cavity unlocked.  When the cavity is locked, light transmitted through the cavity is sent to a polarimeter consisting of a rotating quarter wave plate, polarizing beamsplitter, and another photodiode (PPD = polarizer photodiode) after passing through a non-polarizing beam splitter (NPBS).  Other diagnostics include the photodiode used for the cavity feedback (RPD=reflected photodiode) and a power meter (PM) to monitor the cavity power~\cite{ComponentLibrary}.
\label{fig:laser_DOCP_schematic}}
\end{center}
\end{figure*}

The Compton polarimeter in Hall A at JLab is a significantly upgraded version of the system reported in Refs.~\cite{HAPPEX:2000rub,Baylac:2002en,Escoffier:2005gi}.  
\begin{figure*}[t]
{\includegraphics*[width=0.9\textwidth]{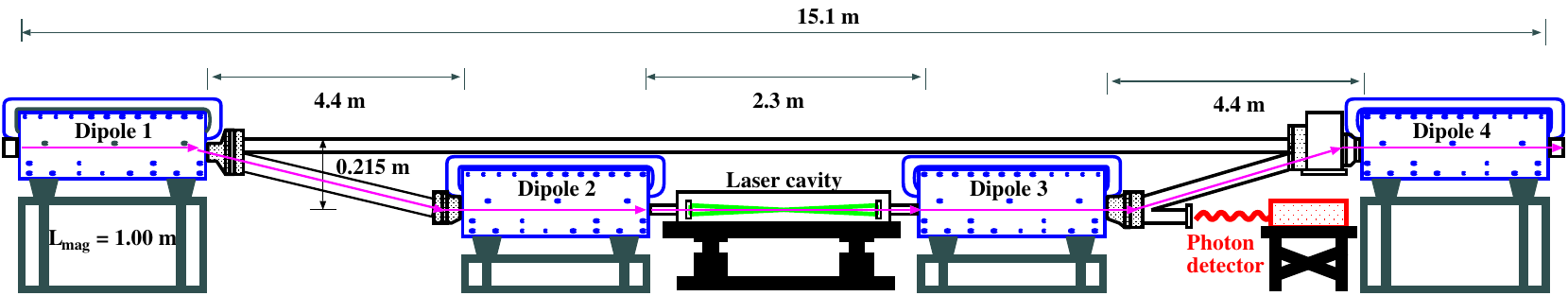}}
\caption{Schematic of the Compton polarimeter in Hall A (not to scale).  The beam is deflected using a magnetic chicane to interact with a laser target.  Compton-scattered photons are detected in an integrating detector.  
Figure adapted from~\cite{Narayan:2015aua}.
\label{fig:compton_layout1}}
\end{figure*}
A schematic of the setup is shown in Fig.~\ref{fig:compton_layout1}.  
The electron beam is diverted vertically in a four-dipole magnetic chicane so it can interact with a photon target.
Circularly polarized green-laser light of 532\,nm wavelength is injected into a Fabry-Pérot optical cavity, in the beam-line vacuum, with a gain of approximately 2500 and a 1.3\degree~crossing angle with the electron beam.
The laser system (details in  Fig.~\ref{fig:laser_DOCP_schematic}) is mounted on a vibration damped optical table between dipoles 2 and 3.
Dipoles 1 and 2 are matched so there is no net polarization precession on the path to the laser interaction point. Similarly, dipoles 3 and 4 return the beam to the original beamline without net precession. 
The Compton-scattered photons pass through an aperture in the third magnet and are detected. The Compton-scattered electrons are momentum analyzed by the third dipole magnet and could be used for polarimetry but were not used in this measurement.  

The Compton analyzing power $A_p$ depends on the energies of the incident and scattered particles.
For an electron beam of 2.18\,GeV, scattering from green (532\,nm) photons, $A_p$ reaches a maximum of 7.5\% at the kinematic endpoint for backscattered photons of 158\,MeV.  Integrated over the full scattered energy spectrum and weighted by the response of the photon calorimeter, the analyzing power is 3.6\%.
For the CREX experiment, the electron beam polarization was held constant for short time ``windows", with new windows of matching or reversed polarization selected at 120\,Hz. These helicity-state windows are generated in ``quartet" patterns of  $+--+$ or $-++-$, with the quartets chosen in a pseudo-random sequence. 
A typical electron beam current of 150\,$\mu$A and cavity laser beam power of 2.2\,kW led to a Compton scattering rate of 210\,kHz. For the measurement techniques used, the asymmetry averaged over the full spectrum could be measured to a statistical precision of 0.5\% of itself in about one hour of continuous data taking, with the primary sources of noise relating to random variations in backgrounds rather than photon-counting statistics. 

The original Hall A Compton polarimeter photon detection and data acquisition system~\cite{Escoffier:2005gi} was upgraded in 2009, with an approach optimized for improved systematic uncertainty at low beam energies~\cite{Friend:2011qh}. The existing lead-tungstate photon calorimeter was replaced with a Ce-doped Gd$_2$SiO$_5$ (GSO) crystal which was sufficiently fast and produced more light.
The data acquisition system was upgraded to support an integrating readout of the photon detector, which eliminates  uncertainties from triggering and threshold effects.
This system was used to measure the polarization for HAPPEX-III~\cite{HAPPEX:2011xlw}, run in 2009, to $dP/P=0.96\%$, dominated by a 0.8\% uncertainty in the laser polarization. 

The polarimeter was further upgraded in 2010 to use a frequency doubled green laser~\cite{Rakhman:2016rsq}, which was critical for the PREX~\cite{Abrahamyan:2012gp} measurement at a beam energy of only 1.06\,GeV. 
The polarization was measured to $dP/P=1.13\%$, and again was dominated by a 0.7\% uncertainty in the laser polarization.
This system in Hall A was also used for d$_2^n$~\cite{JeffersonLabHallA:2014gzr,JeffersonLabHallA:2014mam} (2009), PVDIS~\cite{PVDIS:2014cmd} (2009) and DVCS~\cite{JeffersonLabHallA:2015dwe} (2010) experiments, which had less stringent requirements for the beam-polarization measurements.

The Qweak Experiment~\cite{Qweak:2018tjf}, which ran from 2010--2012 in Hall C at Jefferson Lab, used a 10\,W, 532\,nm laser with 200-fold cavity gain and detected the scattered electron instead of the photon.  
The laser polarization was measured to 0.18\% using an optical reversibility theorem~\cite{Vansteenkiste:93} which allowed a measurement of the polarization to $dP/P=0.59\%$ at an energy of only 1.16\,GeV~\cite{Narayan:2015aua}.  
The uncertainty was dominated by knowledge of the detector efficiency due to unexpected noise from the electron detector and corresponding high thresholds.  
This result brought polarimetry at JLab into the realm of the 1994-1995 run of the SLD experiment at SLAC, which reported a polarization accuracy of $dP/P=0.5\%$ ~\cite{SLD:2000leq,Woods:1996nz} at a significantly higher beam energy 45.6\,GeV.  

In this work, we combine the integrating photon detector used in Hall A with the laser advancements from Hall C to achieve the highest accuracy electron beam polarimetry that we are aware of, and identify areas with further room for improvement. 

\section{Laser system\label{sec:laser}}

The laser system for the Hall A Compton polarimeter has been comprehensively described in Ref.~\cite{Rakhman:2016rsq}.  Here we provide a brief summary.  Green laser light at 532\,nm is provided via a frequency doubled 1064\,nm laser system.  A narrow linewidth NPRO laser (at 1064\,nm) is amplified to 5--7\,W using a fiber amplifier and frequency doubled via a 50\,mm long periodically poled lithium niobate (PPLN) crystal, resulting in a laser power at 532\,nm of about 0.6--1\,W.  The resulting narrow linewidth light is then coupled into a high finesse ($\approx 1.2\times10^{4}$) Fabry-P\'{e}rot cavity.  The stored power in the cavity (typically 2--2.4\,kW) then provides the ``photon target'' for the electron beam.

A crucial component of the Compton laser is the system for preparing and determining the laser polarization inside the Fabry-P\'{e}rot cavity.  
Previous experiments in Hall A had inferred the polarization in the cavity by measuring the polarization in the exit line. This necessitated the use of a transfer function to describe the
the evolution of the laser polarization after the second cavity mirror, as it is transported outside the beamline vacuum via steering mirrors and vacuum exit window.
This technique has the drawback that the transfer function must be determined with the system at atmospheric pressure and with certain beamline elements removed.  Hence, any change in the vacuum window birefringence due to changes in mechanical stress and vacuum pressure were not accounted for.  These effects are potentially significant and must be controlled.   In order to achieve high accuracy, the birefringence within the cavity caused by multiple reflections from the cavity mirrors must also be taken into account.  While the effect of this is small for a single reflection, the cumulative effect for the stored laser light can become significant.  Previous measurements did not consider this effect because they either used a low gain cavity or because the effect was expected to be small compared to the uncertainty with which the laser polarization could be determined.
Here we use the same optical reversibility theorem as was previously used in Hall C, Ref.~\cite{Vansteenkiste:93}, which shows that on reflection from a mirror, the reflected laser beam can be described using the inverse of the matrix of the forward propagating beam. 
As a consequence, starting with a known polarization before the beamline vacuum and cavity and characterizing the returning polarization state allows determination of the polarization state at the first mirror of the cavity without requiring detailed knowledge of the birefringent properties of the optical elements between the initial laser beam and the first cavity mirror.

Determination of the laser polarization for the CREX experiment was performed in three stages (see Fig.~\ref{fig:laser_DOCP_schematic}):
\begin{enumerate}
 \item{With the Fabry-P\'{e}rot cavity at 1 atm and the beamline open, a model of the evolution of the laser polarization (the ``entrance function") from the polarization-defining polarizing beam splitter  (PBS) to the first cavity mirror was constructed by scanning over the full laser polarization phase space using a quarter-wave plate and half-wave plate placed immediately after the PBS, and monitoring the light reflected back from the cavity (when not locked) in the entrance function photo-diode (EFPD),  which collects the light that passes through the PBS in the reverse direction.  
 This technique has been described in Ref.~\cite{Narayan:2015aua}, in which it was used primarily to determine the  quarter and half-wave plate settings that would result in 100\% degree of circular polarization (DOCP) at the cavity entrance.  In this case, we employ the entrance function, to prepare an arbitrary laser polarization state at the entrance to the cavity.}
 \item{To determine the impact of the possible birefringence in the cavity, several measurements were made of the laser polarization after the cavity with the cavity locked. Modulo transmission through the second cavity mirror, this represents the polarization inside the cavity. Measurements were made for a variety of laser polarization states at the input of the cavity, these polarization states being determined by the entrance function measured in the previous step.  Due to limitations of the locking technique and electronics, it was not possible to sample the full region of laser polarization phase space.  Nonetheless, it was possible to determine the cavity birefringence better than 10\%.  The polarization of the light exiting the cavity was measured using a PBS and rotating quarter-wave plate.  A non-polarizing beamsplitter (NPBS) was used to divert 50\% of the beam power from the nominal laser path to this laser polarimeter.  The NPBS had some small birefringence, which was measured prior to the cavity measurements.}
 \item{Once the cavity birefringence had been determined, the beamline was re-assembled and vacuum restored.  The entrance function was measured once again (as in step 1) since the birefringence of the entrance had likely changed. Note that this is a strength of the back-reflection technique in that it can be employed with the cavity under vacuum.  With the updated entrance function and knowledge of the cavity birefringence, the polarization inside the cavity was fully determined.}
\end{enumerate}

\begin{figure}[htb]
{\includegraphics[width=\columnwidth]{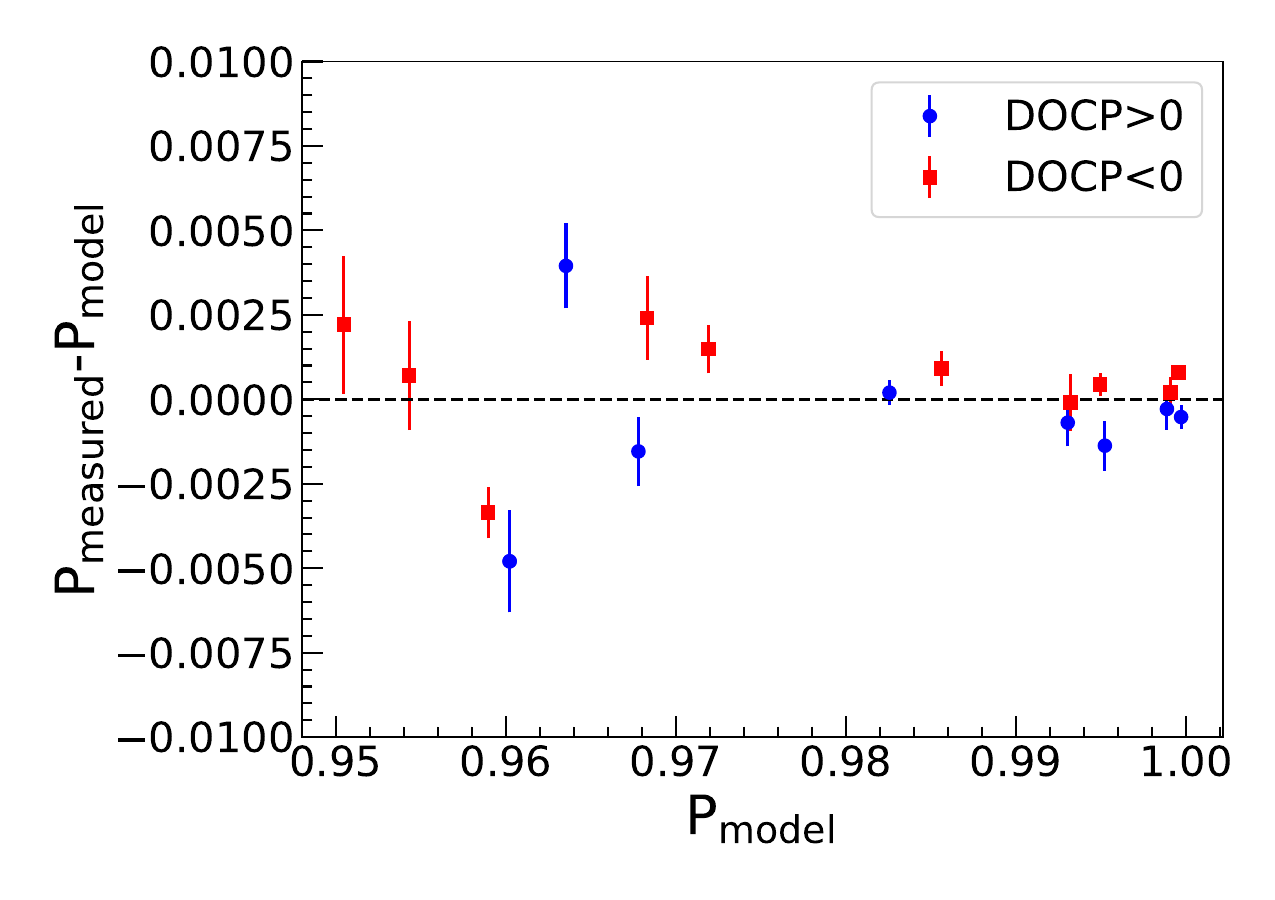}}
\caption{Residuals ($P_\text{measured} - P_\text{model}$) for the fit to the polarization in the Fabry-P\'{e}rot cavity plotted vs. model polarization.
\label{fig:laser_DOCP_res}}
\end{figure}

\begin{figure}[htb]
{\includegraphics[width=\columnwidth]{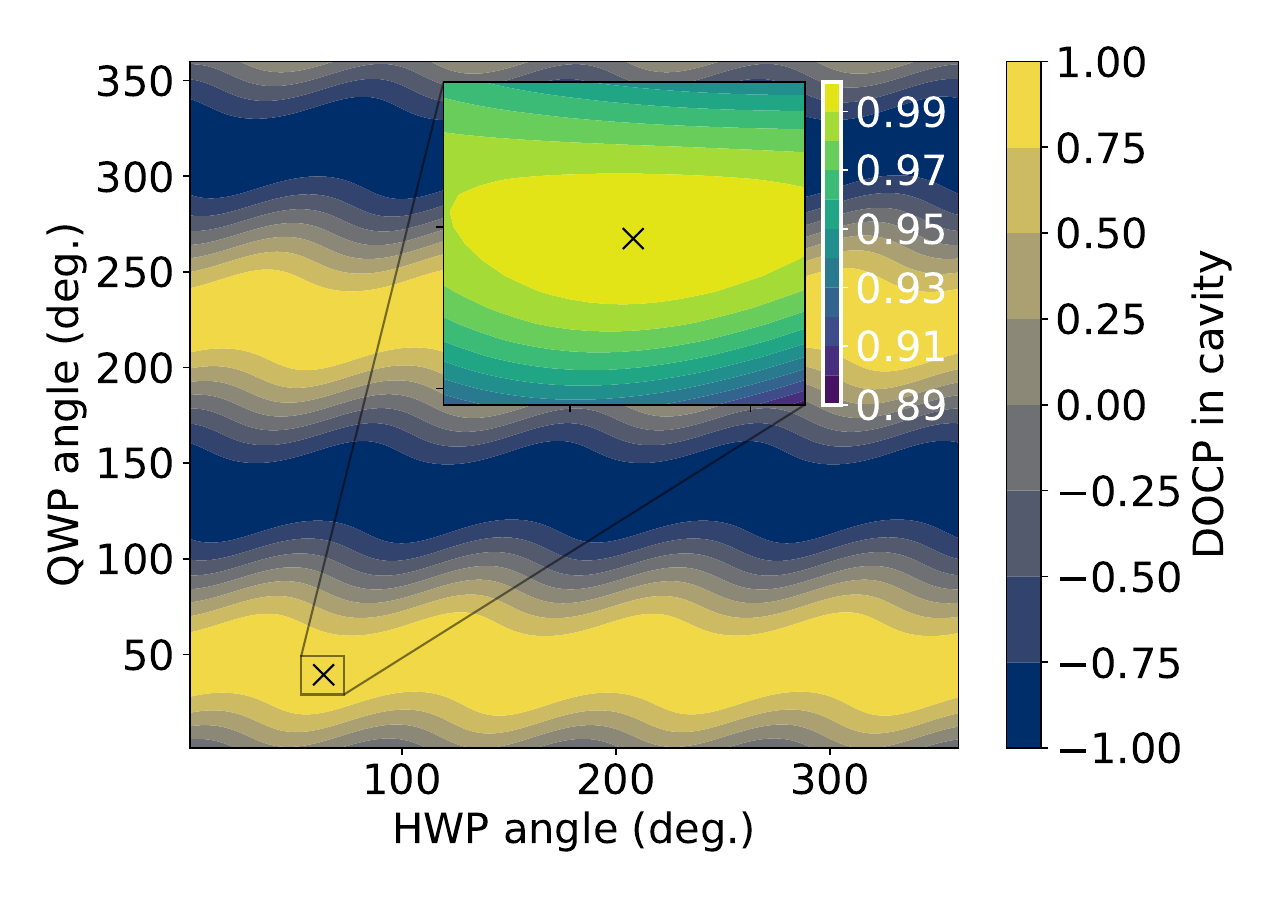}}
\caption{Degree of circular polarization in cavity (model) vs. HWP and QWP angle.  The ``x" denotes the HWP and QWP angles at which the majority of the Compton polarimeter data were taken.
\label{fig:laser_DOCP_2D}}
\end{figure}

During the CREX experiment, the bulk of the Compton polarimeter data was taken with $\theta_\text{QWP} = 39.3^{\circ}$ and $\theta_\text{HWP} = 63.5^{\circ}$, resulting in a degree of circular polarization inside the cavity of $99.99 +0.01/-0.25\%$.  The primary contributions to the uncertainty in the laser polarization are: i) 0.05\% from the observed time dependence of the polarization (monitored by a passive polarimeter outside the cavity), ii) 0.03\% due to uncertainties in the entrance function and cavity birefringence fit parameters, iii) 0.1\% due to possible birefringent effects in the transmission through the second cavity mirror (constrained by direct measurement), and iv) 0.22\% due to the residuals of the model that describes the polarization inside the cavity.  
The latter are shown in Fig.~\ref{fig:laser_DOCP_res}.  
We take the root mean square of the residuals for the region $P_\text{cavity} > 95.0$\% as an indication of the uncertainty due to the fitting technique. The residuals are likely driven by the entrance function, which is extremely sensitive to the laser alignment since small differences in the laser trajectory between the forward-going and back-reflected beam could result in small changes in the birefringence experienced by the laser.  Figure~\ref{fig:laser_DOCP_2D} shows the model calculation of the circular polarization of the laser inside the cavity as a function of quarter-wave and half-wave plate angles.

We expect the laser polarization measurements can be improved by placing the polarization analyzing optics (the PBS, quarter-wave plate (QWP), and half-wave plate (HWP)) closer to the Fabry-P\'{e}rot cavity reducing effects due to non-overlapping incident and reflected laser beam trajectories.  In addition, by using the ``power-balanced detection scheme" implemented in Ref.~\cite{Asenbaum:11} we could capture all the reflected light from the cavity.  This would also allow the cavity to be locked with arbitrary polarization allowing a much greater range of systematic studies.

Note that in an early version of these Compton polarimeter results used in Ref.~\cite{CREX:2022kgg}, a systematic uncertainty of 0.45\% was applied for the contribution of the laser polarization to the overall Compton polarimeter uncertainty.  This larger uncertainty came primarily from the fact that the back-reflection technique allows for two solutions for the entrance function, which in turn allows two solutions for the Fabry-P\'{e}rot cavity birefringence.  The most generic description of a birefringent optical element requires three degrees of freedom (two rotations and one phase)~\cite{Hurwitz:41}. However, the birefringence of a Fabry-P\'{e}rot cavity with two identical mirrors can be expressed using the same form as a generic wave plate (with only two degrees of freedom---one rotation and a phase)~\cite{Moriwaki, biesla}.  Use of the more restricted expression for the cavity birefringence would have allowed selection of a single solution for the cavity birefringence. We chose to employ the more generic prescription, allowing for the possibility that the cavity mirrors were not in fact identical.  Test measurements were taken during the CREX experiment which, after analysis improvements applied after the initial publication of the CREX results, allowed unambiguous determination of the correct entrance function solution (and cavity birefringence) resulting in the reduced uncertainty quoted here.  The corresponding cavity birefringence parameters resulting from the physical entrance function solution turn out to be consistent with a generic wave plate description (with only two degrees of freedom), giving further confidence in the result. 

\begin{figure*}[htb!]
\begin{center}
{\includegraphics*[width=\textwidth]{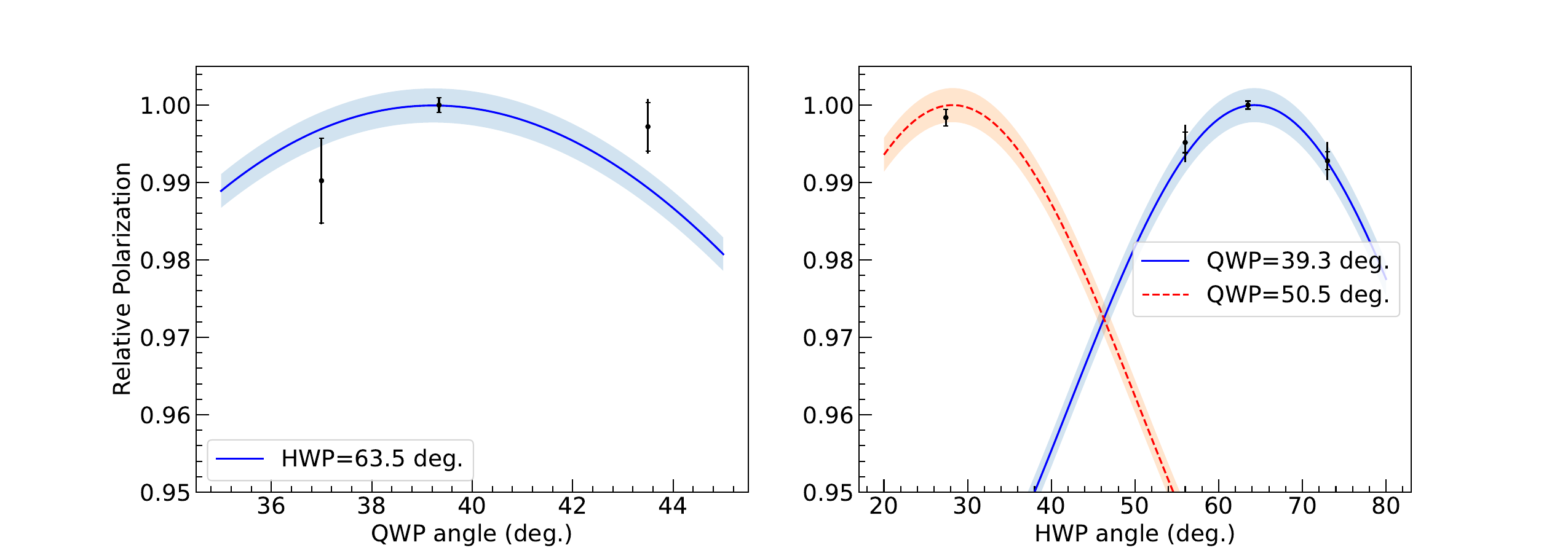}}
\caption{Measurements of the beam polarization with modification to the degree of circular laser polarization.  The figure on the left shows measurements with fixed half wave plate (HWP) angle and varying quarter-wave plate (QWP) angle, while the figure on the right shows measurements vs. HWP angle for two QWP settings.  The beam polarization measurements have been normalized to 1.0 for the nominal QWP/HWP setting of 39.3/63.5 degrees.  The inner error bars on the points show the statistical errors, increased by a factor of $\sqrt{1.3}$ to account for the slightly non-statistical behavior observed in the data (see Sec.~\ref{sec:apower}).  The outer error bar shows the statistical uncertainty combined in quadrature with the uncertainties in the laser polarization due to fitting of the birefringence parameters (generally larger for smaller laser polarization).  The curves show the predictions for the model of the laser polarization, with the shaded bands indicating the 0.22\% fluctuation suggested by the fit residuals.
\label{fig:laserpol_tests}}
\end{center}
\end{figure*}

The results of the test measurements are shown in Fig.~\ref{fig:laserpol_tests}.  A series of Compton polarimeter runs were taken with the laser polarization deliberately changed from its nominal setting in order to test the model of the laser polarization by comparing to the change in the measured Compton asymmetry.  Analysis of these data was complicated by the fact that the electron beam polarization displayed some systematic time dependence for a subset of the runs.  The later analysis removed the time dependence via a fit to the data that did not include the runs with modified laser polarization.  After removal of this time dependence, the laser model solution shown in Fig.~\ref{fig:laserpol_tests} was clearly preferred.

The effective total phase retardation induced by the Fabry-P\'{e}rot cavity was determined as part of the cavity birefringence measurement and was found to be $\delta_\text{eff}$\,=\,1.11\,$\pm$\,0.10\,deg.  Measurements of this quantity have been performed for other cavities and are typically expressed in terms of the phase retardation for a single reflection from the mirrors which is given by $\delta_\text{cav} = \frac{\pi}{2 F} \delta_\text{meas}$, where $F$ is the cavity finesse. For the cavity in this work, $\delta_\text{cav} = (14.5 \pm 1.3) \times 10^{-5}$\,deg, comparable to Fabry-P\'{e}rot cavities using mirrors with similar reflectivity~\cite{biesla} ($R\approx99.98\%$).

\section{Photon detector}
\label{sec:gso}

The detector for the Compton-scattered photons is a cylindrical cerium-doped Gd$_2$SiO$_5$ (GSO) crystal scintillator.  The crystal is  6\,cm in diameter and 15\,cm long (10.9 radiation lengths). 
GSO was chosen for its short pulse duration and relatively high light yield for Compton photons from GeV scale electrons, up to  
158\,MeV for CREX~\cite{Melcher:1990gy, GSOLightYield}. Effects from long duration light ``afterglow" were shown to be negligible compared to a CREX integration window at 120\,Hz. 
The GSO scintillator is mounted on a motorized table which can be remotely moved horizontally and vertically to center the detector on or remove it from the photon flux.
Attached flush to the end of the GSO crystal is a photomultiplier tube (PMT) which collects the scintillation light from the GSO and passes the signal to the data acquisition system~\cite{Friend:2011qh}. 

Upstream of the photon detector is a cylindrical lead collimator to reduce backgrounds from non-Compton processes. This collimator has an outer diameter of 8\,cm, a thickness of 6\,cm and a fixed aperture diameter of 2\,cm and sits approximately 10\,cm upstream of the photon detector with a fixed position. A 2\,cm-diameter disk of 250\,$\mu$m-thick lead is mounted on the front of the photon detector in order to absorb synchrotron radiation from the electron beam.

The position of the photon detector relative to the collimator was determined using two millimeter-thick tungsten ``fingers'', one horizontal placed 2\,cm above, and one vertical placed 2\,cm to the right of the  central axis of the photon detector.  Each finger includes a small scintillator attached behind to measure the rate from the Compton photons converting in the tungsten.
By scanning the detector table vertically or horizontally, the profile of the photon rate can be determined and the detector centered on the maximum.

Asymmetries from the Compton photon detector are formed using a thresholdless, energy-integrating technique,
\begin{equation}
    A_\text{meas}=\frac{\Sigma^+ -\Sigma^-}{\Sigma^+ + \Sigma^-},
\end{equation}
where $\Sigma^{\pm}$ is the total energy of Compton-scattered photons, as measured in the photon detector for the $+$ or $-$ beam helicity during a helicity quartet.  This is equal to the average analyzing power of Compton scattering multiplied by the polarization of both the electron $P_e$ and photon $P_\gamma$, $A_\text{meas} = \langle A_p \rangle P_e P_\gamma$. Ideally, the analyzing power is an energy-weighted average calculated over the full energy spectrum of scattered photons
\begin{equation}
    \langle A_p \rangle_\text{ideal} = \frac{\int_0^{k_\gamma^\text{max}} A_p(k_\gamma) \, k_\gamma \,  \sigma_0(k_\gamma) \, dk_\gamma }{ \int_0^{k_\gamma^\text{max}} k_\gamma \,  \sigma_0(k_\gamma) \, dk_\gamma } 
\end{equation}
where $\sigma_0$ is the unpolarized cross section as a function of scattered photon energy $k_\gamma$. Experimentally, this quantity must be corrected for the acceptance $\epsilon(k_\gamma)$ and average response of the calorimeter $R(k_\gamma)$,
\begin{equation}
\label{eq:AP_meas}
    \langle A_p \rangle_\text{meas} = \frac{\int_0^{ k_\gamma^\text{max}} A_p(k_\gamma) \, k_\gamma\, \epsilon(k_\gamma) \, R(k_\gamma) \,  \sigma_0(k_\gamma) \, dk_\gamma }
    {\int_0^{k_\gamma^\text{max}} k_\gamma\, \epsilon(k_\gamma) \, R(k_\gamma) \, \sigma_0(k_\gamma) \, dk_\gamma} 
\end{equation}
In practice, the photon spectrum was integrated with an upper limit of pulse size corresponding to approximately four times the maximum energy of a Compton-scattered photon, which allowed for linear response measurements even in rare cases of pile-up. The total integral of the photon detector signal was accumulated over each beam helicity window in a flash ADC (fADC) which sampled with 12\,bit precision at 200\,MHz. 

In addition to the required ``integrating mode" used for determination of the Compton asymmetries, the data acquisition (DAQ) simultaneously operates in ``counting mode'' which is used for detector diagnostics, rate calculations, and for obtaining the energy spectrum of detected photons. In this mode, pulse integrals are calculated for a limited sample of pulses, triggered by a constant fraction discriminator using a copy of the photon detector signal. This distribution can be compared to the known Compton-scattering energy spectrum and models of the detector response, for systematic studies.   Energy spectra are available for every polarization measurement and additional,  dedicated runs were intermittently taken with the physics target out of beam (to reduce background) and at higher PMT gain to make higher precision measurements of the Compton energy spectrum.

\subsection{Compton asymmetry calculation}

Backgrounds from the beam must be subtracted  when calculating the asymmetry.
The background is estimated by frequently taking data with the laser off (Fabry-P\'{e}rot cavity unlocked).
The laser system was ``cycled" through on and off approximately every two minutes by automatically locking and unlocking the laser cavity into and out of resonance.
The photon detector data was analyzed in ``laser cycles'' containing one period of laser-on data, and the adjacent periods of laser-off data.

\begin{figure}[htb]
\begin{center}
{\includegraphics[width=\columnwidth]{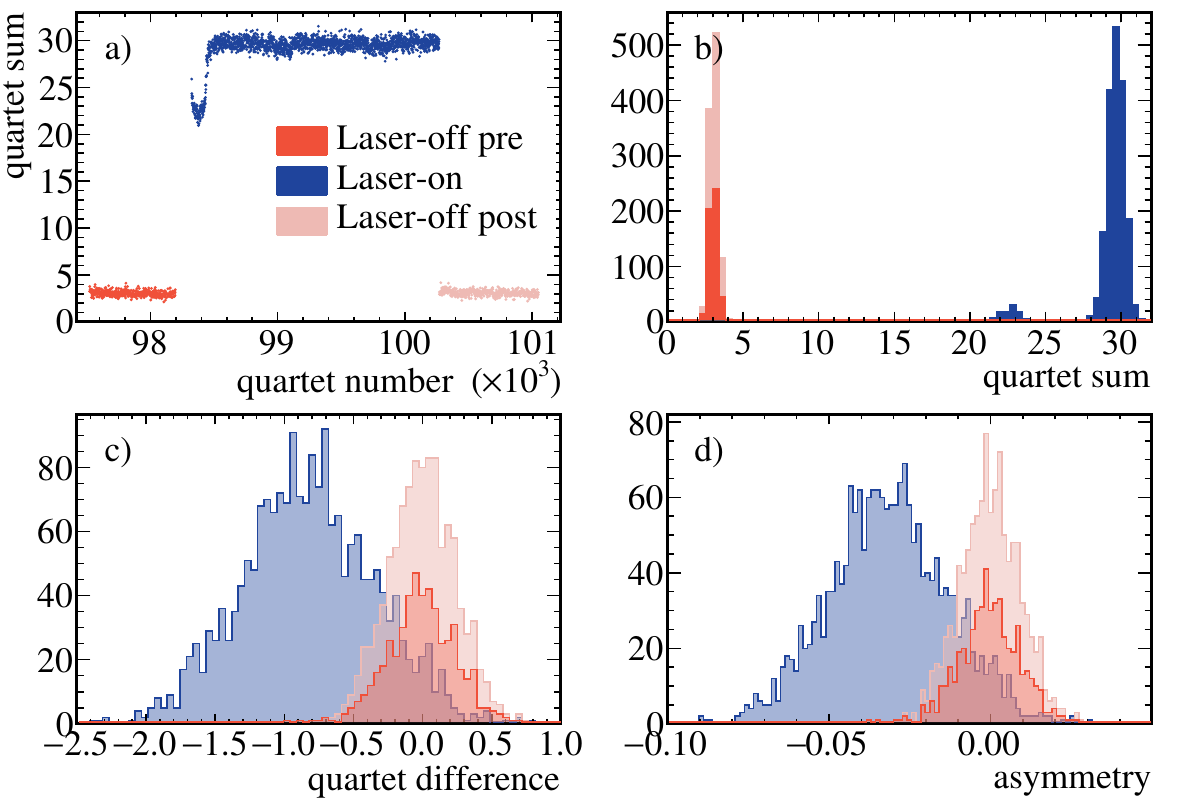}}
\caption{Histograms for quartet quantities in a typical laser cycle.  The events are divided into periods of ``laser-on" and ``laser-off", indicated by color.
a) $Y_\text{ON}$ and $Y_\text{OFF}$, the photon detector yield (sum of photon detector signals over four windows of the helicity quartet) as function of time.  Observable in the  laser-on period is a signal fluctuation due to a laser instability.
b) Histogram of detector yield. 
c) $\Delta_\text{ON}$ and $\Delta_\text{OFF}$, the quartet helicity difference of photon detector signal for laser-on and laser-off periods.  The difference is consistent with 0 for laser-off.
d) $A_\text{ON}$ and $A_\text{OFF}$, the asymmetries calculated as described in the text.
\label{fig:photondetector_lasercycles}}
\end{center}
\end{figure}

The helicity-correlated differences are constructed for the yield in each helicity pattern ($\Delta_\text{ON}$ and $\Delta_\text{OFF}$ for laser-on and laser-off periods, respectively) as well as the  total yield sum ($Y_\text{ON}$ and $Y_\text{OFF}$ for laser-on and laser-off periods, respectively.)
All of these quantities are pedestal subtracted using pedestal values determined during frequent electron beam-off periods.
The Compton asymmetries are then calculated for each helicity quartet  in both laser states as
\begin{align}
    A_\text{ON} &= \frac{\Delta_\text{ON}}{Y_\text{ON} - \langle Y_\text{OFF} \rangle}, \\ 
    A_\text{OFF} &= \frac{\Delta_\text{OFF}}{\langle Y_\text{ON} \rangle - \langle Y_\text{OFF} \rangle},
\end{align}
with the experimental Compton asymmetry for a laser cycle being
\begin{equation}
    A_\text{exp} = \langle A_\text{ON} \rangle - \langle A_\text{OFF} \rangle,
\end{equation}
where the angle brackets $\langle\rangle$ denote an average over the full laser cycle under consideration. 
The helicity difference, yield, and asymmetry data can be seen plotted for a typical laser cycle in Fig. \ref{fig:photondetector_lasercycles}.

\subsection{Determination of the analyzing power\label{sec:apower}}

The analyzing power is determined using a Monte Carlo simulation incorporating realistic photon flux, collimator and detector.   This performs the integral in Eq.~(\ref{eq:AP_meas}) including the energy dependent analyzing power and cross section known from quantum electrodynamics (QED) including radiative corrections~\cite{Denner:1998nk,Zec:2022ida} with the product of acceptance and response,  $\epsilon(k_\gamma)R(k_\gamma)$, from the detector model.

\begin{figure}[htb]
\centering
\includegraphics[width=\columnwidth]{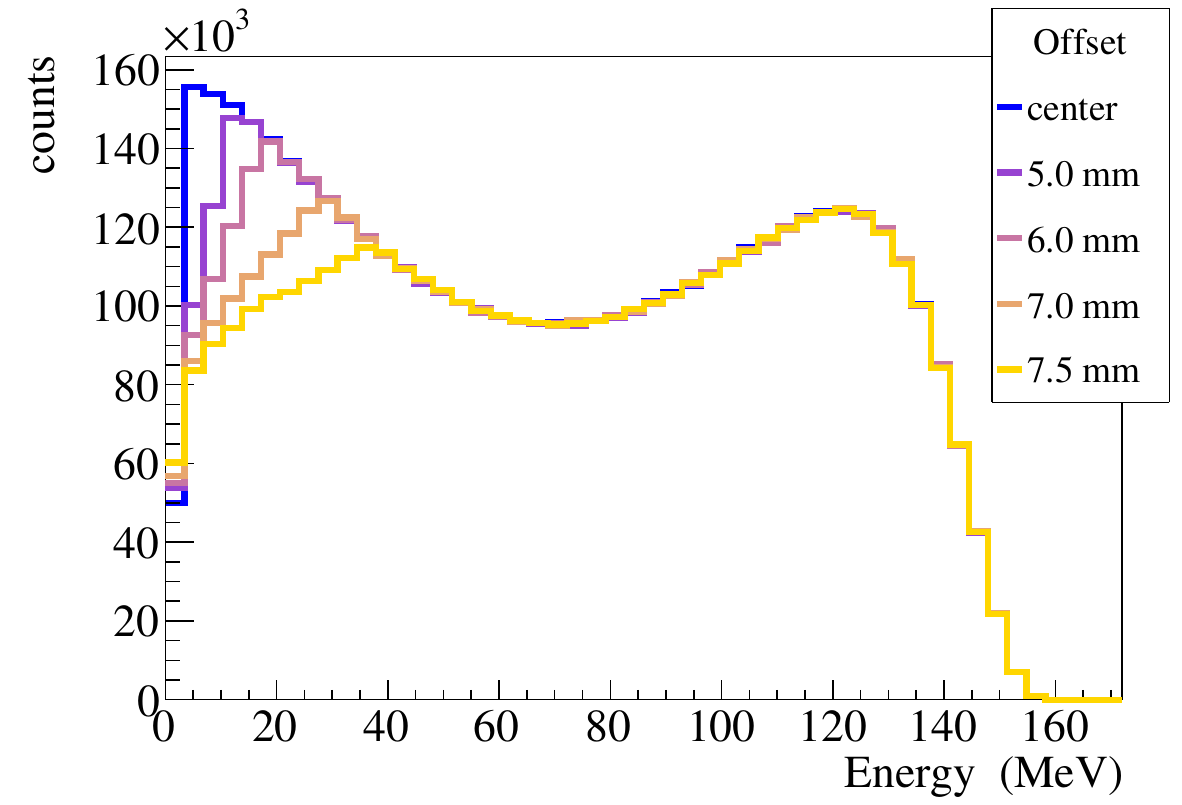}
\caption{Simulation of the spectrum of Compton photons in GSO detector for different amounts of offset between the photon flux and the detector collimator.  Offsets cause lower energy photons with larger production angles to be absorbed, leading to a distortion of the spectrum.  See text for details. 
}
\label{fig:compton_spectra_with_offset}
\end{figure}

The experimental analyzing power depends on the alignment of the photon flux with the collimator aperture.
If the central axis of the cone of scattered photons is offset from the central axis of the collimator then lower energy photons with larger production angles may be absorbed in the collimator, leading to a distortion of the spectrum and an increase in the analyzing power.
The effect on the Compton spectrum as this offset increases can be seen in simulation results in Fig. \ref{fig:compton_spectra_with_offset}. 
This effect was observed during CREX, with the spectra from some runs showing this distortion.
By comparing the measured spectral shape with the Monte Carlo spectrum the size of the offset and the effect on $A_p$ could be estimated.

\begin{figure}[htb]
\centering
\includegraphics[width=\columnwidth]{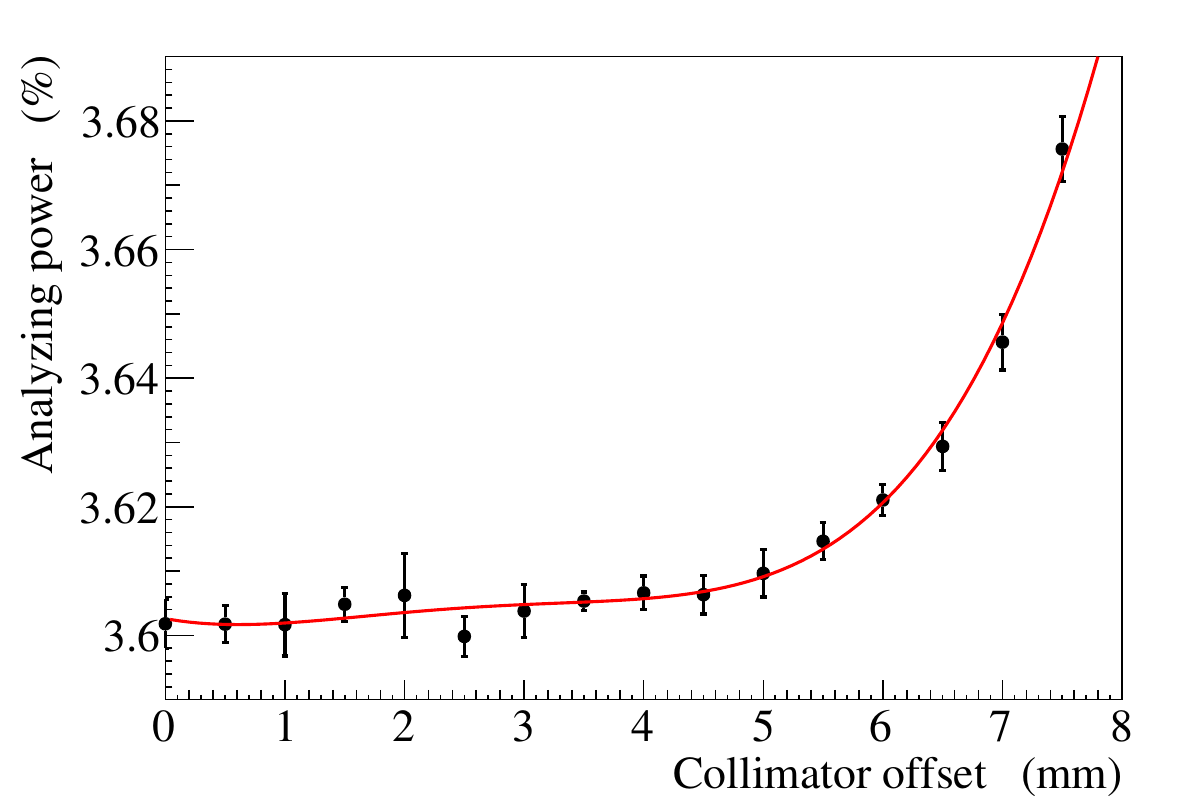}
\caption{Average experimental analyzing power $\langle A_p \rangle_\text{meas}$ as a function of offset between photon flux and collimator, as determined by simulation. The curve is a simple polynomial fit to the analyzing power. 
}
\label{fig:analyzingpower_vs_offset}
\end{figure}

Figure~\ref{fig:analyzingpower_vs_offset} shows the analyzing power as a function of the photon flux offset.
The change in analyzing power is negligible until the offset exceeds 5 mm, beyond which it increases rapidly. 
The beam position and trajectory, from which the back-scattered photon trajectory could in principle be calculated, is measured on the laser table using beam position monitors (BPM) upstream and downstream of the Compton interaction point. 
In practice, this projection showed some inconsistency with the  offsets estimated from spectra.
We believe that this inconsistency is caused by a slow variation in the laser table height, with respect to the fixed collimator and photon detector, in response to changes in the atmospheric pressure interacting with the air cushion in the isolation legs. 
A feedback mechanism keeps the electron beam position constant with respect to the BPMs, which are attached to the table, by adjusting the electric current in the chicane dipole magnets. 

Additional studies of the relative photon-collimator offset used the rapid variation in analyzing power at large offsets, relative to the high statistical precision of the polarimeter, to independently bound a possible average collimator offset.  It is noted that a significant average position offset would also necessary imply large changes in $A_p$ for small variations in the offset consistent with expected beam position variations. These studies showed that the statistical consistency of the polarimeter data set over long timescales ($\chi^2/\nu=1.3$, Fig.~\ref{fig:compton_results_w_moller}) can rule out an average collimator offset large enough to produce a shift of $\delta A_p / A_P > 0.2\%$.  This bound, larger than the corrections implied by either the collimator centering calibrations or the BPM pointing during production running, was adopted as a limit of systematic uncertainty.

\subsection{LED pulser\label{sec:LED}}

The integrating measurement technique has the benefit of removing the sensitivity of the experimental $A_p$ to knowledge of the absolute energy calibration of the GSO+PMT system.  However, the measurement depends crucially on the linearity of the detector response over the range of pulse sizes and pulse rates. 

The PMT linearity can be tested \textit{in situ} using a system of pulsed light emitting diodes (LEDs) built-in to the photon detector housing~\cite{Friend:2011pw}. 
This ``LED pulser'' system works by flashing two LEDs in a repeating sequence with a frequency of 250 Hz. 
The flashing sequence has four parts: both LEDs flash simultaneously, each LED flashes individually, and then both LEDs remaining off.
One of the two LEDs (``Variable'') is allowed to vary, decreasing in brightness as the sequence progresses, while the other (``Delta") stays at a fixed brightness.  Once started, the flashing sequence is controlled by an automated circuit on the DAQ. The LED pulser sweeps the Variable LED from its maximum value above the Compton edge brightness down to zero, and the PMT light yield is recorded for each part of the LED sequence. 

\begin{figure}[htb]
    \begin{center}
        {\includegraphics[width=\columnwidth]{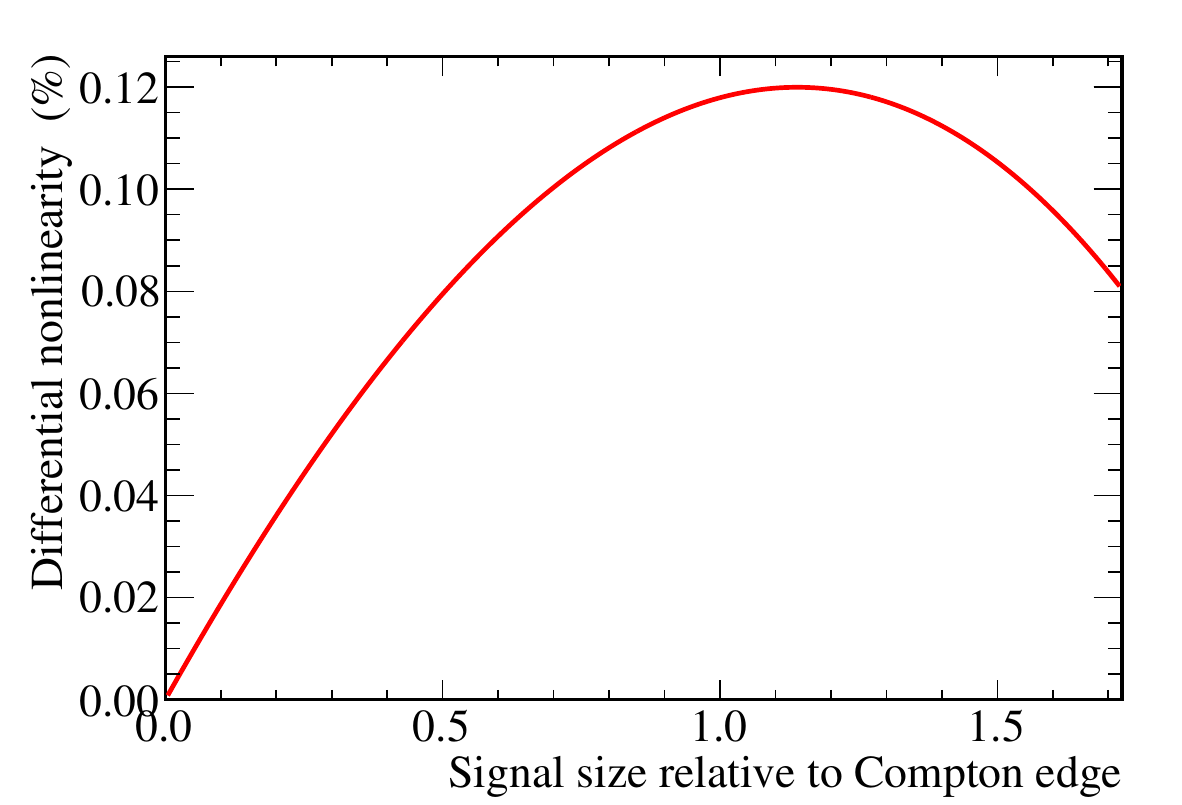}}
        \caption{Photon detector PMT nonlinearity function. For this plot, the signal size is normalized to the average signal size of a photon at the Compton edge.
        \label{fig:nonlinearity}}
    \end{center}
\end{figure}

The finite-difference linearity is given by 
\begin{equation}
    \varepsilon = \frac{Y(V+\Delta)-Y(V)}{Y(\Delta)}
\end{equation}
where $Y(V)$, $Y(\Delta)$ and $Y(V+\Delta)$ are the yields for the Variable, Delta and simultaneous flashing respectively.
For a perfectly linear PMT, the finite-difference linearity function would be exactly 1 over the measured energy range.  The yield of the detector is parametrized as a fourth order polynomial, with the parameters fit to the measured finite-difference nonlinearity.  The resulting differential nonlinearity is shown in Fig.~\ref{fig:nonlinearity}. 
The nonlinearity is used to apply a correction in the simulation, increasing the measured energy by up to 0.12\%, depending on the energy. The nonlinearity of the PMT was found to contribute a 0.02\% uncertainty to the polarization measurement.
A ``Dark Delta'' LED outside the PMT housing is used to study potential cross-talk between the LEDs, which would invalidate the measurement principle.  None was found.

There is a third (``Load'') LED in the PMT housing which shines at a  constant brightness allowing the replication of various loads.
This is used to characterize the PMT ``gain shift'', that is, a change in the PMT gain as a function of PMT rate or total brightness. 
The most significant effect of such a gain shift would be in the subtraction of the background signal, due to the large variation in average illumination between laser-on and laser-off periods. 
The PMT gain shift was characterized by measuring the pulse height with a constant LED brightness for loads corresponding to laser-on and laser-off running. 
The gain shift $\alpha$ was defined as:
\begin{equation}
    \alpha = \frac{Y^{\Delta}_\text{ON} - Y^{\Delta}_\text{OFF}}{Y^{\Delta}_\text{OFF}},
\end{equation}
where $Y^{\Delta}_\text{ON}$ is the reference pulse signal height with an average load matching 
$\langle Y_\text{ON}\rangle $ 
and $Y^{\Delta}_\text{OFF}$ is the reference signal pulse height with the average signal matching $\langle Y_\text{OFF}\rangle$. 
Given $\alpha$, a correction can be applied to the Compton asymmetry as
\begin{equation}\label{eq:gainShiftCalc}
    \langle A_\text{corr} \rangle = \frac{\langle A_\text{exp} \rangle + \alpha f \Delta_\text{OFF}}{1 + \alpha f Y_\text{OFF}},
\end{equation}
where
\begin{equation}
    f = \frac{1}{Y_\text{ON} - Y_\text{OFF}}.
\end{equation}

This technique was applied to a similar PMT prior to the experiment and found to be $\alpha=0.001$.
During the CREX experiment the system failed and beam data had to be used to determine an upper bound of $\alpha<0.012$.  Through Eq.~(\ref{eq:gainShiftCalc}) this corresponds to a maximum relative change in asymmetry of 0.15\%. No correction to the asymmetry was applied and the maximum bound was taken to be the uncertainty.

\section{Results}\label{sec:results}

In total, the CREX Compton data set contained 15,232 laser cycles, 14,498 of which passed data quality cuts on pedestal stability, minimum signal size, minimum statistical power, consistent laser-off asymmetry, and small charge-asymmetry.

\begin{figure}[htb]
\begin{center}
{\includegraphics[width=\columnwidth]{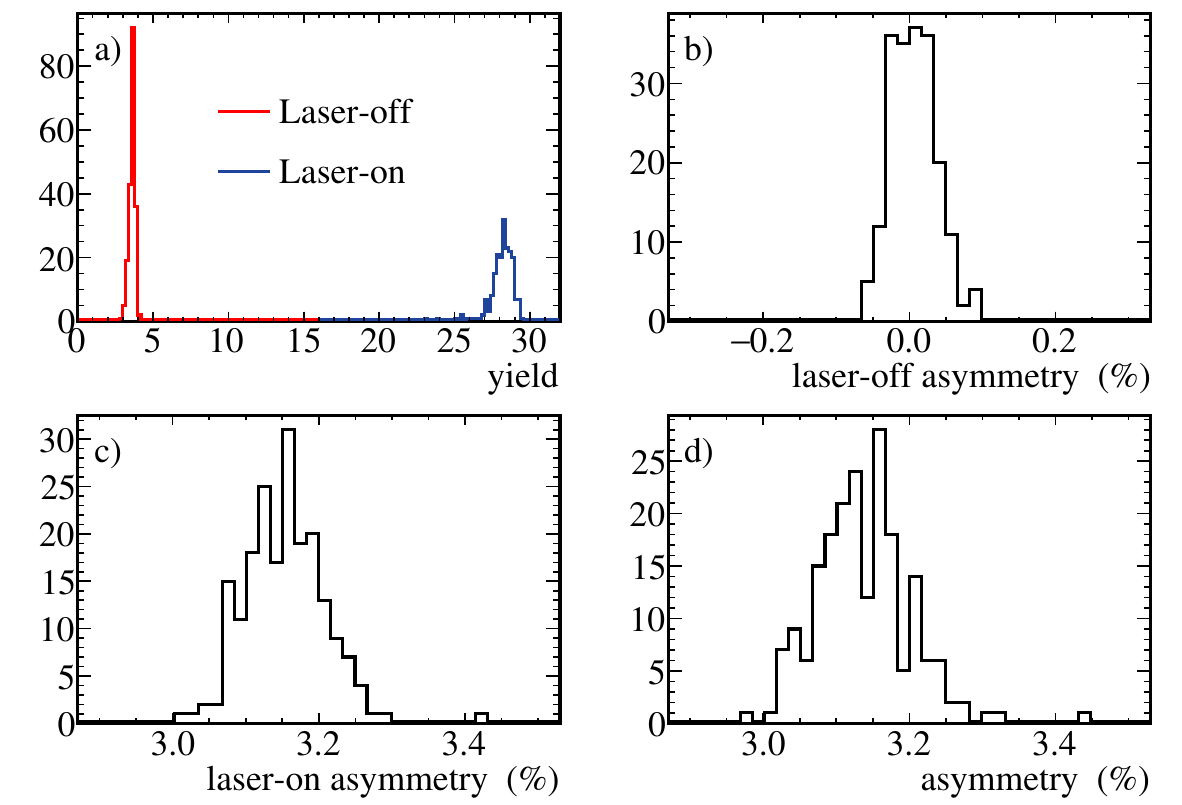}}
\caption{Histograms for the laser cycles in a typical snail.  a) $\langle Y_\text{OFF} \rangle$ and $\langle Y_\text{ON} \rangle$, the pedestal subtracted yield for laser-on and laser-off,   b) $\langle A_\text{OFF} \rangle$, the  asymmetry for the laser-off period.   c) $\langle A_\text{ON} \rangle$, the asymmetry for the laser-on period.  d) $A_\text{exp}$, the extracted experimental asymmetry for the cycle.
\label{fig:cycles_in_snail_asym}}
\end{center}
\end{figure}

\begin{figure}[htb]
\begin{center}
{\includegraphics[width=\columnwidth]{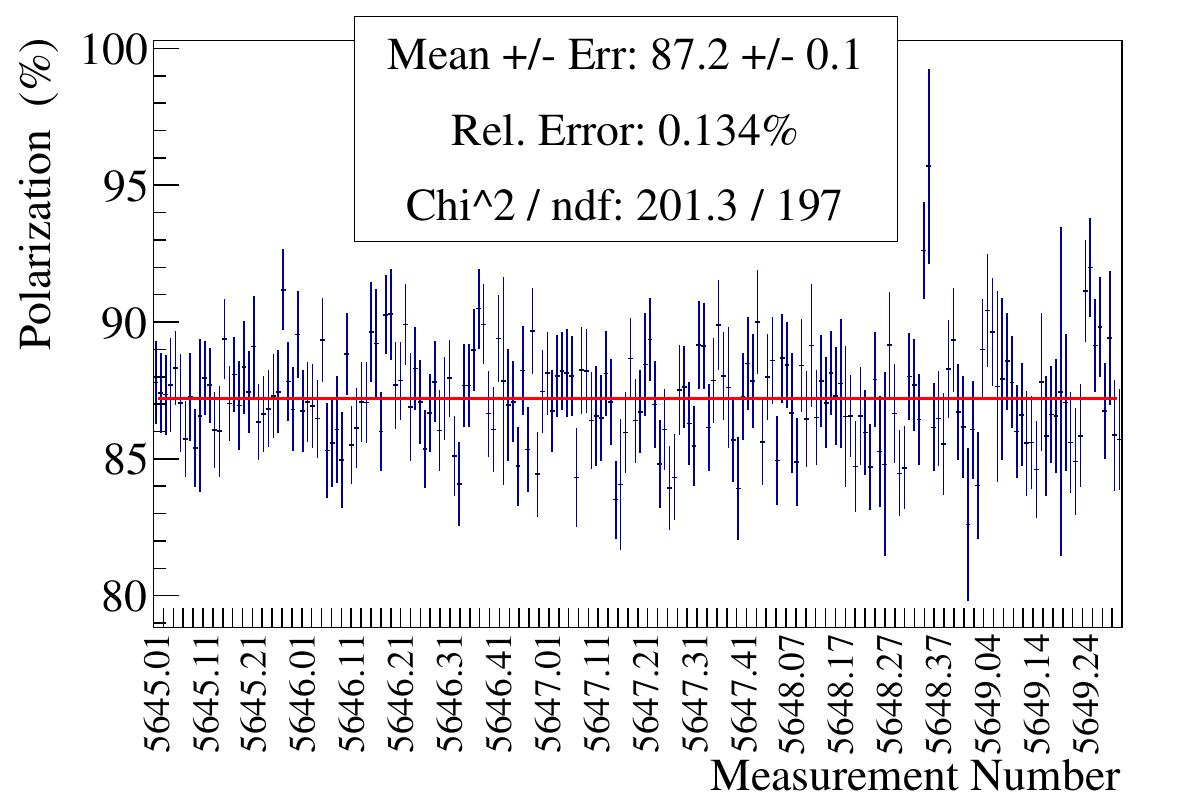}}
{\caption{Polarization of each laser cycle in a typical ``snail'' ($\approx8$\,h period) plotted versus cycle number.  Uncertainties are statistical.  The combined polarization, shown in the text box, is the uncertainty-weighted mean of the cycle polarizations.}
\label{fig:cycles_in_snail}}
\end{center}
\end{figure}

During experimental running, the relative polarization direction of the beam was flipped periodically as a means of controlling sources of systematic uncertainty for the CREX experiment. 
This flip would also change the sign of the polarization, and thus had to be analyzed separately with a sign correction applied for the final polarization analysis. 
These periods, known colloquially amongst the collaboration as ``snails''\footnote{The term ``snail" corresponds roughly with data periods called ``slugs" in the CREX experiment, hence the name.} provided an aggregated measurement of polarization over the span of approximately 8\,h throughout the experiment.
The average polarization for each snail was calculated as the uncertainty-weighted average of the measured laser cycle polarizations contained within that snail. 
The average yields and asymmetries for each cycle in a typical snail are plotted in Fig.~\ref{fig:cycles_in_snail_asym}.
The polarization measurement for a typical snail can be seen in Fig. \ref{fig:cycles_in_snail}.

\begin{figure*}[htb]
{\includegraphics*[width=\linewidth]{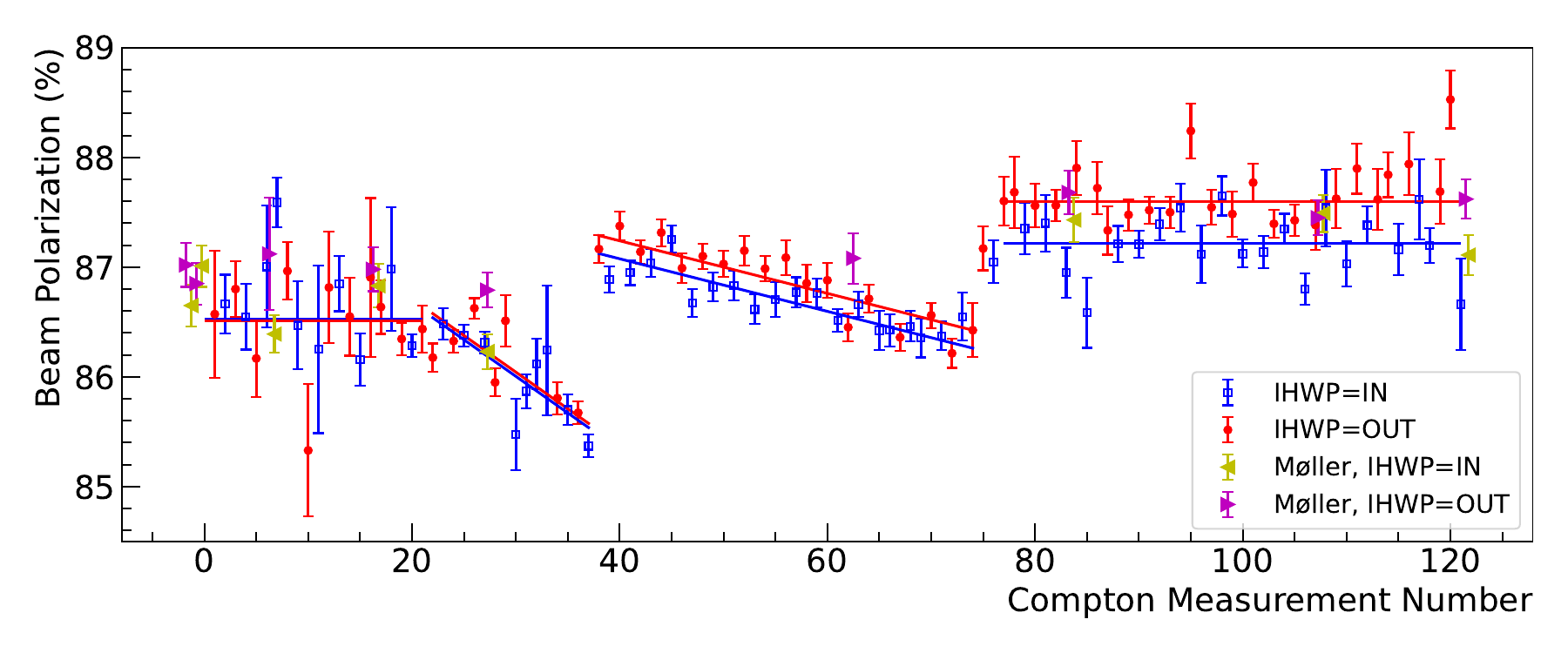}}
\caption{Measured beam polarization during the CREX experiment from the Compton polarimeter.  Uncertainties are statistical only.  The observed changes in polarization are due to changes in the quantum efficiency of the photocathode which increases when ``reactivated" and may decay slowly with time.  The data is consistent with the  fit shown with $\chi^2=1.3$.  The difference in polarization between the two states of the insertable half-wave plate is due to unaccounted for birefringence in the vacuum window in the polarized source.  Polarization measurements from the M\o ller polarimeter are also shown.  The two polarimeters are in excellent agreement.
\label{fig:compton_results_w_moller}}
\end{figure*}

The polarization of the Jefferson Lab electron beam is known to vary slowly with time due to the dependence of the polarization on the quantum efficiency of the photocathode in the polarized beam source~\cite{10.1063/1.4994306}.
This was also observed during CREX, as can be seen in Fig.~\ref{fig:compton_results_w_moller}.

This changing polarization may cause a dependence on the timescales at which the polarization correction is applied to the asymmetry data. 
The average polarization for the experiment was determined by aggregating the data over various timescales, including averaging over the full run, snail-by-snail corrections, or interpolating a smoothed fit for each $A_{PV}$ production slug. Results were found to vary by no more than 0.02\%,  
which was assigned as the uncertainty due to the averaging timescale.

Systematic uncertainties not discussed earlier include that from absolute beam energy and helicity correlated beam positions and unequal polarization in the two helicity states.
The measurement of CREX beam energy was reported with a 0.05\% relative uncertainty, corresponding to a 0.05\% uncertainty in $A_p$.

An unequal polarization in the two electron helicity states leads to a correction to the inferred polarization proportional to the size of the analyzing power multiplied by half the difference in the polarization~\cite{Friend:2011qh}.  Measurements with the polarized electron source bound the difference in DOCP of the laser at the photocathode (and hence the electron beam polarization) to be $<1.2\%$.  This leads to a maximum correction of $0.03\%$, which is used as the uncertainty.

The measurement is very insensitive to position differences.  The electron beam waist at the interaction point decreases the size of the position differences.  Average position differences over the run are consistent with zero.  The electron beam position is locked to BPMs of the laser table to maximize the overlap of the electron and laser beams, which minimizes sensitivity to position differences. 
The correction for the measured position differences averaged to $<0.01\%$ over the whole run.  A value of $0.01\%$ was taken as the uncertainty.

\begin{table}[ht]
    \centering
    \begin{tabular}{lcc} 
    \hline
    Source                           &     $\sfrac{dP}{P}$(\%)\\ 
    \hline
    Laser polarization               &     0.25    \\
    Collimated spectrum distortion   &     0.20  \\
    Detector gain shift              &     0.15   \\
    Beam energy                      &     0.05   \\
    Helicity state polarization difference    &     0.03   \\
    Detector nonlinearity            &     0.02   \\
    Averaging timescale              &     0.02   \\
    Position differences              &     0.01   \\
    \textbf{Total}                   & \textbf{0.36}  \\ \hline
    \end{tabular}
    \caption{The final systematic uncertainties are dominated by a knowledge of the laser polarization. See text for details.}
    \label{tab:compton_systematics}
\end{table}
The systematic uncertainties for the CREX Compton measurement are summarized in Table \ref{tab:compton_systematics}, leading to a total systematic uncertainty on the Compton measurement of $dP/P=0.36$\%.  
The Compton polarimetry result, 86.90 $\pm$ 0.31\% (syst) $\pm$ 0.02\% (stat), is an average over the polarization measurements shown in Fig.~\ref{fig:compton_results_w_moller} weighted by the square-inverse of the statistical uncertainty in the parity-violating asymmetry in the main CREX measurement taken in the same time period.  In this way, the quoted polarization and uncertainty best reflects the polarization normalization for the APV measurement.

In addition to the Compton polarimeter, the CREX experiment made use of a M\o ller polarimeter to provide a second measurement of the beam polarization with independent systematic uncertainties (0.85\% relative)~\cite{King:2022auk}.
The beam polarization results from both the Compton and M\o ller polarimeters are shown in Fig.~\ref{fig:compton_results_w_moller}.  
Each Compton measurement represents one ``snail" (discussed earlier) which is the error weighted average over (typically) several hours of data.
The M\o ller measurements have a residual compared to the Compton fit of $0.29\pm0.05\%$~(stat).
The results are consistent between polarimeters given the systematic uncertainties.

\subsection{Future developments}

Future parity-violation experiments in Hall A, such as  MOLLER~\cite{MOLLER:2014iki} (a measurement of $A_{PV}$ in elastic electron-electron scattering) and PVDIS~\cite{JeffersonLabSoLID:2022iod} (measurements of $A_{PV}$ in deep inelastic electron scattering), will require electron beam polarimetry with a relative precision of $0.4\%$.  
These future experiments will be run at beam energies of 11~GeV and 6.6~GeV, which provide substantially larger asymmetries and a peak energy for back-scattered photons that is a larger fraction of the beam energy, compared to the CREX measurement.  In this way, the higher beam energies should be expected to provide similar or improved control of systematic uncertainties in Compton polarimetry.
The addition of an electron detector in the Hall A Compton polarimeter will be particularly useful at these higher energies. 
As noted above, there are also further improvements which can be made in the determination of the laser polarization. 
In addition, it will be important to minimize slow drifts of the laser table and implement a method of tracking the table position.  This may be particularly important at higher beam energies where smaller collimating apertures may be needed to reduce the impact of synchrotron radiation.

\section{Conclusion}

The polarization of the electron beam was continuously measured during the running of the CREX experiment via Compton polarimetry to an accuracy of $dP/P=0.36$\%, reducing the impact of this uncertainty below other significant contributions to the total uncertainty in the CREX result.
This result achieves, for the first time, the uncertainty required for the high profile future parity-violation experiments in Hall A.

\begin{acknowledgments}
The authors would like to thank Dipangkar Dutta (Mississippi State University) for providing the source upon which Fig~\ref{fig:compton_layout1} is based. This material is based upon work supported by the U.S. Department of Energy, Office of Science, Office of
Nuclear Physics under Contracts No. DE-AC05-06OR23177, DE-FG02-87ER40315, and DE-FG02-07ER41522.
\end{acknowledgments}

\bibliography{references}

\end{document}